\documentclass[runningheads]{llncs}
	\pdfoutput=1
	\usepackage{graphicx}
	\usepackage{float}
	\usepackage{mathtools}
	\usepackage{xcolor}
\usepackage[utf8]{inputenc} 
\usepackage[T1]{fontenc}    
\usepackage{lmodern}
	\usepackage{algorithm}     

	\usepackage[hidelinks]{hyperref}
	\usepackage{algpseudocode}   
	\usepackage{dcolumn}        
	\usepackage{bm}              
	\usepackage{amsmath,amssymb}

	\DeclareMathOperator{\cc}{gcc}
	\newcommand{\bigO}{\mathcal{O}}          
	
	\DeclareMathOperator{\diam}{diam}
	\DeclareMathOperator{\ecc}{ecc}

\begin{document}
	\title{Constrained graph generation:\\ Preserving diameter and clustering coefficient simultaneously}

	\titlerunning{Constrained graph generation}

	\author{Dávid Ferenczi \inst{1}\orcidID{0009-0000-9507-5774},
	Alexander Grigoriev\inst{1}\orcidID{0000-0002-8391-235X}}

	\authorrunning{D.~Ferenczi, A.~Grigoriev}

	\institute{Maastricht University, Maastricht, 6211LM, Netherlands
	\email{\{d.ferenczi,a.grigoriev\}@maastrichtuniversity.nl}}

	\maketitle              
		\begin{abstract}
 Generating graphs subject to strict structural constraints is a fundamental computational challenge in network science. Simultaneously preserving interacting properties---such as the diameter and the clustering coefficient--- is particularly demanding. Simple constructive algorithms often fail to locate vanishingly small sets of feasible graphs, while traditional Markov-chain Monte Carlo (MCMC) samplers suffer from severe ergodicity breaking. In this paper, we propose a two-step hybrid framework combining Ant Colony Optimization (ACO) and MCMC sampling. First, we design a layered ACO heuristic to perform a guided global search, effectively locating valid graphs with prescribed diameter and clustering coefficient. Second, we use these ACO-designed graphs as structurally distinct seed states for an MCMC rewiring algorithm. We evaluate this framework across a wide range of graph edge densities and varying diameter-clustering-coefficient constraint regimes. Using the spectral distance of the normalized Laplacian to quantify structural diversity of the resulting graphs, our experiments reveal a sharp contrast between the methods. Standard MCMC samplers remain rigidly trapped in an isolated subset of feasible graphs around their initial seeds. Conversely, our hybrid ACO-MCMC approach successfully bridges disconnected configuration landscapes, generating a vastly richer and structurally diverse set of valid graphs.
	\keywords{Graph generation  \and ACO \and MCMC}
	\end{abstract}

		\section{Introduction}
		
		Generating graphs with prescribed properties is a well-known problem at the intersection of random graph theory~\cite{bollobas_2001,molloy_critical_point} and network science~\cite{newman_networks,maslov2002specificity}; however, the core question in these two fields is stated differently. From a theoretical perspective, inquiry focuses on constructive existence: Can a member of a graph family $\mathcal G$ be generated in polynomial time? From a practical point of view the question is of statistical significance: If we were to generate a random ensemble of graphs satisfying a given set of constraints, how likely the resulting set of graphs maintain the required features? To address the second question one would need an efficient and unbiased sampling method for random constrained graph generation. However, if the constraints are too strict, even theoretical constructive algorithms might fail to find valid graphs, either due to the low cardinality of $\mathcal G$ compared to the number of possible graphs (vanishingly small feasible region), or due to computationally costly checks needed to verify the properties (computational complexity). In case of properties that can be checked in polynomial time, Markov-chain Monte Carlo (MCMC) methods can be used~\cite{robert2004monte_carlo}, but these methods also might fail: MCMC methods might not be able to generate a representative sample of $\mathcal G$ due to multimodality~\cite{brooks2011handbook}, therefore introduce bias towards certain instances of $\mathcal G$.
		
		To address this challenge we turn to using Ant Colony Optimization (ACO) algorithms~\cite{blum_ACO,dorigo_ACO} for their ability to explore $\mathcal G$ efficiently by constructing graphs one edge at a time. Using ACO-generated graphs as diverse seeds for MCMC, we can overcome the MCMC multimodality issue. Our two step framework combines using ACO to explore $\mathcal G$ by finding various seed graphs, and then using MCMC for sampling, yet preserving the desired properties of generated graphs.

		In this paper we want to sample graphs with their number of nodes and edges fixed, such that the diameter and clustering coefficient are kept within given parametrized bounds. Notice, we are not defining a variant of the configuration model--- a random graph model with a fixed degree distribution--- as we do not intend to keep the degree sequence or distribution intact, though our methods can be tailored to cover such a case as well. 
		
		The challenge lies in the nature of the two restricted parameters, as they strongly interact with each other: increasing clustering coefficient creates structural redundancy decreasing the diameter of the graph \cite{Watts1998,Newman2003_Structure}. While specific generative models have been proposed to maintain these two properties \cite{Holme2002}, sampling of constrained graph ensembles remains a computationally difficult issue.
		Another challenge comes from the fact that the diameter is a global property of a graph~\cite{van2024random,Newman2003_Structure}: the effect of the displacement of an edge on the diameter is unpredictable.
        
        \begin{figure}[htbp]
			\centering
			\includegraphics[width=\linewidth]{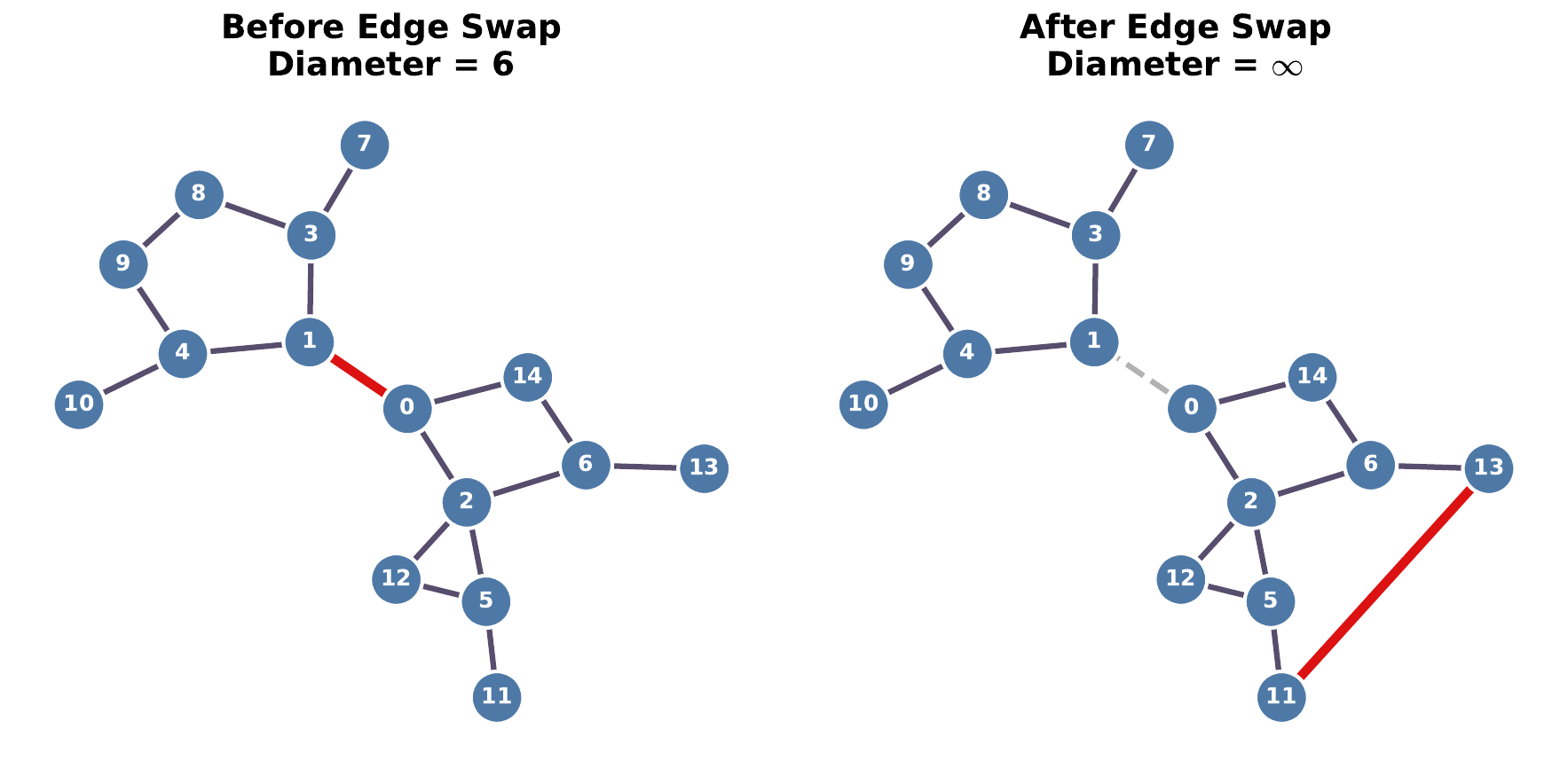}
			\caption{\label{fig:diam_collapse} Impact of an edge swap on diameter. Moving a single red edge from a central position (Left) to a boundary position (Right) fundamentally alters the diameter.}
		\end{figure}
		
        \section{Problem statement and preliminaries}\label{sec:problem}
		
		We start with definitions, and a brief overview of the algorithms which will be used further in the paper.
		
		Let $G=(V,E)$ be a simple graph on $n$ nodes, and $m$ edges. A \textit{triplet} in a graph is 3 nodes connected by two edges, and a \textit{closed triplet} is a graph on $3$ nodes, that forms a triangle. With these definitions at hand, we can define the \textit{global clustering coefficient} if the number of triplets is non zero as:
 \begin{equation*}
				\cc(G)=\frac{ \text{number of closed triplets}}{\text{ number of all triplets}}
\end{equation*}
and define $\cc(G)=0$, if the number of all triplets is zero. The number of closed triplets in a simple graph is 3 times the number of triangles in the graph \cite{Hofstad_2016}. With this we can rewrite the definition above as \begin{equation*}
	\cc(G)=\frac{3\cdot\text{number of triangles}}{\text{number of triplets}}.
\end{equation*} We will use the notation $\Delta_G$ and $\Lambda_G$ for the number of triangles, and triplets respectively. This quantity is often also regarded in the literature as \emph{transitivity}, or \emph{global clustering coefficient}.
\begin{lemma}[\cite{Hofstad_2016}]\label{lem:clustering_trace}
			Let $G$ be a graph with degree sequence $(\delta_i)_{i=1}^n$, and $A$ its adjacency matrix. Then \begin{equation*}
				\cc(G)=\frac{\frac{1}{2}\cdot tr (A^3)}{\sum_i^n \binom{\delta_i}{2} }.
			\end{equation*}
\end{lemma}

The \textit{diameter} of $G$ is $\diam(G) = \max_{u, v \in V} d(u,v)$ and the \textit{eccentricity} of a node $u\in V$  is $\ecc(u)=\max_{v\in V\setminus \{u\}} d(u,v)$, where $d(u,v)$ is the length of the shortest path connecting $u,v$. Using the eccentricity we can state the following well-known bound for the diameter.
		\begin{lemma}[\cite{Hofstad_2016}]\label{lem:diameter_eccentricity}
			Let $G=(V,E)$, then \begin{equation*}
				\ecc(v)\leq \diam(G)\leq 2\cdot \ecc(v),
			\end{equation*}
			for any node $v\in V$.
		\end{lemma}
		
Now we formally state the problem. Given the number of nodes and edges $n,m$, and intervals $[\cc_{\min},\cc_{\max}]$, $[\diam_{\min},\diam_{\max}]$, we want to generate a set of graphs $G_i$, such that $\cc(G_i)\in [\cc_{\min},\cc_{\max}]$ and $\diam(G_i) \in [\diam_{\min},\diam_{\max}]$.
		
		\subsection{Simple edge swaps}
		We continue by describing the core operation we do while manipulating the graphs. An \textit{edge swap} is the deletion and insertion of an edge between two connected nodes $(u,v)$ and $(x,y)$, that is often used for graph rewiring algorithms \cite{Alstott_rewiring}. We use the notation $(u,v)\to (x,y)$ for swapping $(u,v)$ to $(x,y)$. 
		
		We can see in Figure $\ref{fig:diam_collapse}$, that the diameter can behave erratically under simple edge swaps. Therefore we will focus on the clustering coefficient first as that is only affected by the local changes in the graph. 
		
		For an arbitrary node $x\in V$, we call the set of nodes $N_x=\{u|(u,x)\in E\}$ the neighborhood of $x$. For the degree of node $x$, we use the notation $\delta_x$.
		A naive recalculation of $\cc(G)$ after every move would cost $\mathcal{O}(n \langle \delta \rangle^2)$, with $\langle \delta \rangle$ being the expected degree in the graph. However, we can achieve $\mathcal{O}(\delta_{max})$ efficiency by observing that changes to the triangle count are strictly local.
		
		\subsubsection*{Incremental update rule for $\cc$.} When an edge is moved from $(u,v)$ to $(x,y)$, the change in $\Delta_G$ is determined exactly by the common neighbors of the active nodes:
		\begin{itemize}
			\item \textbf{Deletion:} Removing $(u,v)$ eliminates exactly $|N_u \cap N_v|$ triangles, and $\delta_u+\delta_v-2$ triplets
			\item \textbf{Insertion:} Adding $(x,y)$ creates exactly $|N_x \cap N_y|$ new triangles and $(\delta_x+\delta_y)$ triplets
		\end{itemize}
		
		Thus, the update logic requires only four neighborhood intersections, bounded by the maximum degree $\delta_{max}$. The number of triplets is updated arithmetically based on the degree changes of $u, v, x,$ and $y$. 
		
		Let $G$ be the graph before the move $(u,v)\to (x,y)$, and $G'$ after. The clustering coefficient $\cc(G')$ can be updated as:
		
		\begin{equation}
			\cc(G') = \frac{3(\Delta_G- |N_u \cap N_v| + |N_x \cap N_y|)}{\Lambda_G - (\delta_u + \delta_v - 2) + (\delta_x + \delta_y)}.
		\end{equation}
		 The numerator adjusts for the elimination and creation of triangles, while the denominator accounts for the change in the number of connected triplets centered at the affected nodes.
		\subsubsection*{Effect on the diameter}
		Due to the global nature of the diameter, there are no guaranteed bounds for the change of the diameter under edge swaps. In order to maintain the diameter within bounds we can turn to one of the following two ways:\begin{itemize}
			\item After each edge swap recalculate the diameter. This can be done using the All-Pair-Shortest-Path (APSP) algorithm to calculate the eccentricity for each node, and by picking the maximum eccentricity we can acquire the diameter. The runtime for APSP is $\bigO(n(n+m))$, and if this check is executed after each edge swap, the runtime of the algorithm is  $\bigO(m(n(n+m)))$ if we want a constant fraction of edges displaced from their original position.
			\item To avoid the computational expense of exact diameter calculation, we employ the Double-Sweep Breadth-First Search (BFS) heuristic \cite{Aingworth_fast_est,Magnien_fast_diam}. This algorithm runs in linear time $\bigO(m)$ and produces an estimator $\hat D$. This value serves as a guaranteed lower bound and, when combined with Lemma \ref{lem:diameter_eccentricity}, provides a strict range for the true diameter: \begin{equation*} \hat{D} \leq \diam(G) \leq 2\hat{D}. \end{equation*} In practice, the lower bound $\hat{D}$ often coincides with the exact diameter\cite{Magnien_fast_diam}. 
		\end{itemize}
		\section{Algorithmic description}
		We continue by providing an overview of our algorithm which is built upon the results from Section \ref{sec:problem}. 
		
		We first describe a Metropolis-Hastings algorithm in subsection \ref{subsec:MH-alg}, that takes a graph as an input and rewires its edges, while keeping $n,m$ fixed, and $\diam$ and $\cc$ within parametrized and previously fixed bounds. This is not a proper graph generation method, as it can not generate graphs from scratch given $n,m,\diam$ and $\cc$ but can rewire, and therefore sample given a seed graph.
		
		To overcome this we describe our ACO algorithm in subsection \ref{subsec:ACO-alg}, which can construct graphs given a list of desired parameters. We later combine these two by running ACO first to get a set of graphs that can be used as an input for the MCMC to reshuffle, and therefore get uniformly distributed samples. We encode graphs as edge lists for better performance.
		\subsection{Metropolis-Hastings algorithm}\label{subsec:MH-alg}
		By proposing edge swaps and accepting the new graphs if the new clustering coefficient and diameter are within an acceptable range, we define a rewiring algorithm that keeps both properties within bounds. 
		
		With this simple mechanism we are in fact describing a Metropolis-Hastings algorithm which can be used to sample uniformly \cite{robert2004monte_carlo,earl2005parallel} at random --- produce an independent sample from the set of constrained graphs--- as long as the corresponding Markov chain is ergodic and the proposal distribution is symmetric (for a brief overview on the Metropolis-Hastings algorithm see Appendix \ref{app:MHalg}). The Metropolis-Hastings step is formally described in Algorithm \ref{alg:mcmc_step}:

		\begin{algorithm}[htbp]
			\caption{Metropolis-Hastings Step}
			\label{alg:mcmc_step}
			\begin{algorithmic}[1]
				\State \text{\textbf{Input}: Given iteration $t$ of the rewiring algorithm:} $(G(t), \Delta(t), \Lambda(t),\cc_{\min},\cc_{max},\diam_{min},\diam_{max})$
				\State \textbf{Output:} Next State $(G(t+1), \Delta(t+1), \Lambda(t+1))$
				
				\Statex
				\State \textbf{1. Proposal:}
				\State Select an existing edge $(u,v)$ and non-edge $(x,y)$ uniformly at random from $E(G)$, and $V\times V \setminus (E(G))$
				\State Define candidate graph $G' \coloneqq (G(t) \setminus \{(u,v)\}) \cup \{(x,y)\}$
				
				\Statex
				\State \textbf{2. Check Local Constraints (Exact):}
				\State Calculate $\Delta', \Lambda'$ and $C'$ using Incremental Update Rule (Eq.~1)
				\If{$C' < C_{min}$ \textbf{or} $C' > C_{max}$}
				\State \Return $(G(t), \Delta(t), \Lambda(t))$ \Comment{Reject: $\cc$ violation}
				\EndIf
				
				\Statex
				\State \textbf{3. Check Global Constraints (Heuristic):}
				\State $D_{est} \gets$ \Call{DoubleSweepBFS}{$G'$}
				\If{$D_{est} > D_{max}$ \textbf{or} $D_{est} < D_{min}$}
				\State \Return $(G(t), \Delta(t), \Lambda(t))$ \Comment{Reject: Diameter violation}
				\EndIf
				
				\Statex
				\State \textbf{4. Accept:}
				\State \Return $(G', \Delta', \Lambda')$
			\end{algorithmic}
		\end{algorithm}
	
\begin{figure}[htbp]
			\centering
			\includegraphics[width=\linewidth]{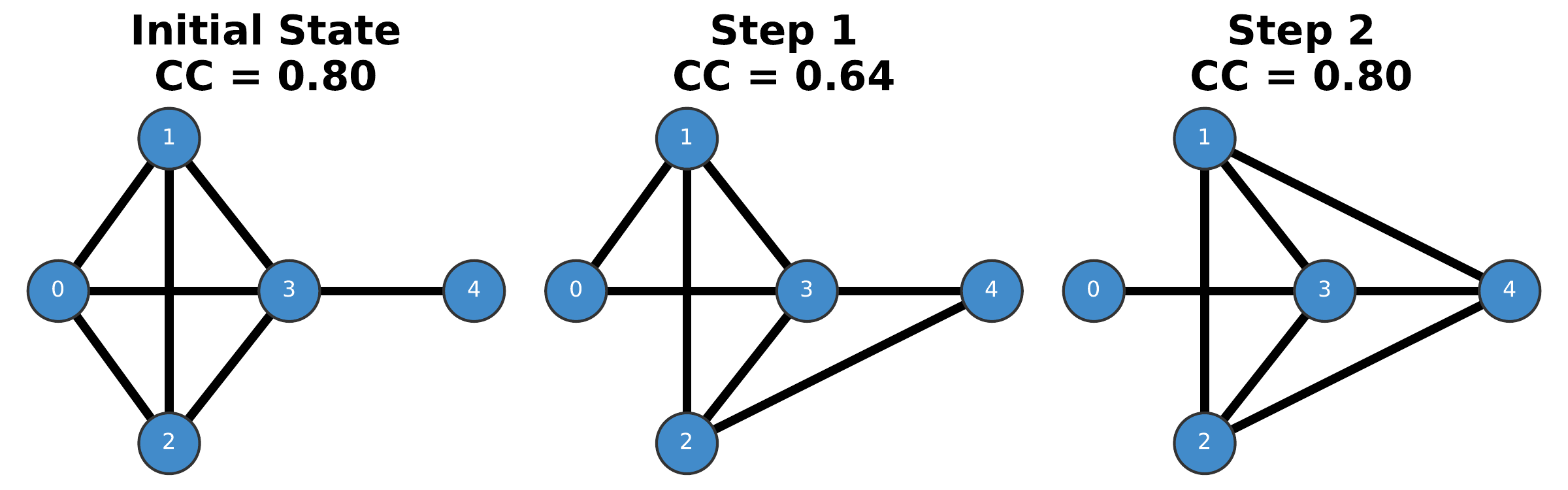}
			\caption{\label{fig:issue_in_statespace} Illustration of the connectivity barrier in the configuration space. To transition the clique from the left side (Step 0, $\cc \approx 0.8$) to the right side (Step 2, $\cc \approx 0.8$), the graph must pass through an intermediate state (Step 1) with significantly lower clustering ($\cc \approx 0.64$). If the constraint is strictly set to $\cc \in [0.7, 0.9]$, the move to Step 1 is rejected, leaving the algorithm trapped in the initial topology.}
		\end{figure}

		In the appendix we show that the proposal distribution is symmetrical. The other property--- ergodicity is not guaranteed. In fact if the bounds for the clustering coefficient are too close to each other the chain will not be able to visit every allowed state. An example of this phenomenon is displayed on Figure \ref{fig:issue_in_statespace}.
		
		Since our chain is not ergodic, we are not able to sample from the set of every allowed graph, only a subset of those graphs. This subset can be described as graphs, that are reachable from $G_0$---the initial starting graph--- by only doing simple and valid edge swaps. This highly limits the versatility of the graphs we are building. To overcome this limitation, we turn to Ant Colony optimization.
		
				\subsection{ACO for Graph Construction}\label{subsec:ACO-alg}
	 	 To resolve the issue explained in Section \ref{subsec:MH-alg} practically, we propose a hybrid framework where Ant Colony Optimization (ACO) serves as a \textit{constructive} global searcher \cite{blum_ACO,dorigo_ACO}. 
	 	 
	 	 We turn to this method, as we want to find as structurally different graphs as possible, while keeping our constraints. What ACO does, is that it starts as a random search, and governs itself in the direction of feasible solutions, by learning from the mistakes and good choices it made while searching. We start building the graph edge by edge, and once we have reached the number of edges needed we check for constraints, and reinforce the choices for certain edges if they brought us closer to a solution. In the ACO literature this step is called pheromone score update, where pheromone scores are deposited for edges that brought us closer to our target. These pheromones on edges mean that the ants will be more likely to pick those while constructing the next graph.

		In a standard ACO framework adding any edge would be allowed. By doing that we would be doing a search, where the vast majority of graphs generated by ants would be instantly rejected, due to the diameter constraint.
		
		To overcome this, and exploit the constructive nature of ACO, we are using a layered constructing heuristic, which will ensure that the diameter remains within bounds.
		\subsubsection{Layered construction}
		In order to guarantee a lower bound $\diam_{\min}$ for the diameter we split the set of nodes $n$ into $\diam_{min}+1$ groups $L_1,L_2,\ldots L_{\diam_{{\min}+1}}$. A layer of nodes is a subset of vertices that constrains the edge formation in order to preserve diameter. Edges are only allowed to form within layers, and between neighboring layers. Formally we create the set $\mathcal E_{available}$, consisting of allowed edges in the graph \begin{equation*}
			(u,v)\in \mathcal E_{available} \iff |\text{layer}(u)-\text{layer}(v)|\leq 1.
		\end{equation*}
		\subsubsection{Pheromones and edge probabilities}
		To continue building our ACO algorithm we need to specify how the pheromone scores affect ants preferences towards certain edges. Let $\tau_{e}$ be the pheromones corresponding to the edge $e=(i,j)$. The probability for an edge being chosen by an ant is given by \begin{equation}\label{eq:probability_pheromones}
			p_{e}=\frac{\tau_{e}}{\sum_{o\in \mathcal{E}_{available }}\tau_{o}}.
		\end{equation}

		 Let $\tau_{e}$ be the pheromones corresponding to the edge $e=(i,j)$. In the beginning, this is uniformly distributed, i.e. \begin{equation*}
			\tau_{e}=\begin{cases}
				1 \text{ if } e\in \mathcal E_{available}\\
				0 \text{ else}.
			\end{cases}
		\end{equation*}

		At each step of the algorithm $k$ ants try to construct a valid graph by picking $M$ distinct edges from $\mathcal E_{available}$ where every edge is chosen according to the distribution described in (\ref{eq:probability_pheromones}).
		
		Once the $k$ ants have constructed the graphs we update 
		the pheromone scores on the edges after evaluating the ants performance by using a reward function. We do so by calculating the realized diameter and clustering coefficient, and defining the reward function corresponding to the $\ell$-th ant's graph as follows:
		\begin{equation*}
			R_k=\begin{cases}
				\frac{\diam_{\text{valid}}}{\varepsilon+|C^\star-\cc(G_k)|}\text{, if }\hat{D}\leq D_{max}\\
				\frac{\diam_{\text{invalid}}}{\varepsilon+|C^\star-\cc(G_k)|}\text{, else}
			\end{cases}
		\end{equation*}
		where $C^\star$ is $\frac{\cc_{\min}+\cc_{\max}}{2}$, $\hat D$ is the estimation of the diameter acquired by doing a double sweep $BFS$, and $\diam_{\text{valid}}$, and $\diam_{\text{invalid}}$ are constants used to reward ants that produce graphs with valid diameters. In our example implementation we use $1$ and $0.1$ for them.
		
		This reward function can be used to update the pheromones on the edges to boost the formation and dissolution of triangles as follows.
		\subsubsection{Pheromone score update}
		
		We employ an elitist pheromone update strategy to make sure that the ants who get closer to valid graphs can steer the others to look for solutions. This means that once all the $k$ graphs have been constructed we rank them based on the graphs reward score, and new pheromones are updated based on the elite ants (i.e. the graphs associated with the highest fitness values).
		
		To guide the search we use the fact that the formation of intra-layer edges boosts triangle generation, whereas the formation of inter-layer edges shrinks the number of triangles. We use the parameter $\rho$ for pheromone evaporation rate. This parameter is responsible for the algorithm's memory to be less restrictive, and ants forgetting about their past.
		
		\begin{equation}
			\tau_{e} \leftarrow (1-\rho)\tau_{e} + \sum_{a \in S_{\text{elite}}} R_a \cdot W_{e}^a.
		\end{equation}
		where $W^a_e$ controls for the formation of triangles. If the clustering coefficient is lower than the target, it raises it, and if the clustering coefficient is higher than the target it lowers it.
		\begin{equation}
			W_{e}^a= 
			\begin{cases} 
				B & \text{if } \cc(G_a) < C^*  \text{ and } e \text{ is intra-layer}, \\
				H & \text{if } \cc(G_a) < C^* \text{ and } e \text{ is inter-layer}, \\
				H & \text{if } \cc(G_a) > C^* \text{ and } e \text{ is intra-layer}, \\
				B & \text{if } \cc(G_a) >  C^*\text{ and } e \text{ is inter-layer},
			\end{cases}
		\end{equation}
		where the parameters $B$ and $H$ correspond to boosting and hindering edge formation within or between layers. In our implementation $B=2$ and $H=0.5$. Algorithm \ref{alg:aco_graph} shows the formal description of our algorithm.

		\begin{algorithm}[htb]
			\caption{ACO for Target Diameter and Clustering Coefficient}
			\label{alg:aco_graph}
			\begin{algorithmic}[1]
				\Require $n,m, \cc_{\min},\cc_{\max}, \diam_{\min},\diam_{\max},T,K$
				\State Initialize $\tau_{ij} \gets 1$ for all valid pairs
				\State Define layers $L_1, \dots, L_{\diam_{\min}+1}$ randomly
				\For{$t = 1$ to $T$}
				\State $Solutions \gets \emptyset$
				\For{$k= 1$ to $K$}
				\State $G_k \gets (V, \emptyset)$
				\State Select $M$ edges probabilistically based on $\tau$
				\State Calculate $\cc(G_k)$ and $\hat{\diam}(G_k)$
				\State Compute fitness $R_k$ based on $\cc(G_k), \hat{\diam}(G_k)$
				\State Save graphs with fitness $(G_k, R_k)$
				\EndFor
				\State Sort saved graphs by fitness
				\State Update Pheromones: $\tau \gets \tau \cdot (1 - \rho)+\sum_{a\in S_{elite}}R_aW_e^a$
				\EndFor
				\State \Return Valid graph found
			\end{algorithmic}
		\end{algorithm}
		
		\subsection{Hybrid Framework: Initializing MCMC with ACO} We propose a hybrid strategy that leverages the complementary strengths of ACO and MCMC approaches: \begin{itemize} 
			
			\item Initialization via Global Search: We first utilize the ACO algorithm to generate a set of distinct, feasible graph instances satisfying the target constraints $(n,m,\cc_{\min},\cc_{\max},\diam_{\min},\diam_{\max})$. Unlike random graph generation with prescribed number of edges, which may fail to locate the feasible region in highly constrained spaces, ACO acts as a guided global search to provide valid starting points. 
			\item Mitigating Ergodicity Constraints: The ACO-generated graphs serve as seed states for the Metropolis-Hastings algorithm. In the space of constrained graphs, the solution landscape is often disconnected, as shown in Figure \ref{fig:issue_in_statespace}. This can cause standard MCMC samplers to become trapped, and not be able to explore every valid graph. By seeding the sampler with topologically diverse solutions from ACO, we ensure the MCMC explores multiple distinct regions of the solution space, thereby generating more representative graphs of the final ensemble.
			
		\end{itemize}
		\section{Experimental results}
In this section we will report some results on our ACO-MCMC hybrid approach. All code and data for reproducing results are available online\footnote{ \url{https://github.com/ferenczid/constrained_graph_generation}}. 

In section \ref{subsec:succ_rat} we report the effect of edge density on the $\textit{success ratio}$--- the rate of successfully constructing graphs by ACO--- and $\textit{structural diversity}$--- a quantity measuring the structural difference between the generated graphs.
 
In \ref{subsec:diff_mcmc} we report the difference in \textit{structural diversity} between graphs generated by MCMC and our ACO-MCMC. 

\subsection{Success ratio and structural diversity}\label{subsec:succ_rat}
In this section we demonstrate the effectiveness of our ACO-MCMC framework for generating structurally diverse graphs.
We evaluated the \emph{success ratio} and \emph{structural diversity} for given pairs of $\diam$ and $\cc$ for a graph on 40 nodes with various density. 

For each pair of parameters 100 instances of ACO were run with the input parameters of the number of nodes, edges, $\cc$ and $\diam$. Each instance had 40 ants looking for solutions. \textit{Success ratio} was calculated as the fraction of trials successfully finding graphs abiding the given constraints. 

To quantify the \textit{structural diversity} of the generated graph ensembles, we use the \textit{spectral distance} of the normalized Laplacian \cite{Wilson_Spectra,Spectral_Gu}, normalized by the number of nodes in the graph. Let $G$ be a graph obtained by our framework, $D$ be the diagonal matrix constructed from the degrees of $G$, and $A$ be the corresponding adjacency matrix. The normalized Laplacian $\mathcal L$ is defined as $\mathcal{L} = I - D^{-1/2} A D^{-1/2}$\cite{chung1997spectral}, and let $\bm{\lambda}(G) = (\lambda_1, \lambda_2, \dots, \lambda_N)$ denote the sequence of its eigenvalues sorted in non-decreasing order ($0 = \lambda_1 \le \dots \le \lambda_N \le 2$). The spectral distance of $G$ and $G'$ is \begin{equation*}
	d_{\text{spectral}}(G,G')=\sqrt{\frac{\sum_{i=1}^N (\lambda_i-\lambda_i')^2}{N}}
\end{equation*} 
For a given set of constraints (target diameter $\diam$ and global clustering coefficient $\cc^*$), the algorithm attempts to generate an ensemble of $L$ valid graphs, denoted as $\mathcal{G} = \{G_1, G_2, \dots, G_L\}$. To quantify the overall functional diversity of this ensemble, we calculate the mean pairwise spectral distance:
\begin{equation*}\mathcal{D}_{\text{ensemble}}(\mathcal{G}) = \frac{\sum_{1 \le i < j \le L} d_{\text{spectral}}(G_i, G_j)}{\binom{L}{2}} .
\label{eq:ensemble_diversity}\end{equation*}
This metric $\mathcal{D}_{\text{ensemble}}$ represents the average structural variation within the solution space located by the algorithm. A lower value indicates graphs that are structurally similar, where high values show there are structurally different graphs satisfying the constraints\cite{Wills_graph_comp}.
\begin{figure}[htb]
	\includegraphics[width=1\columnwidth]{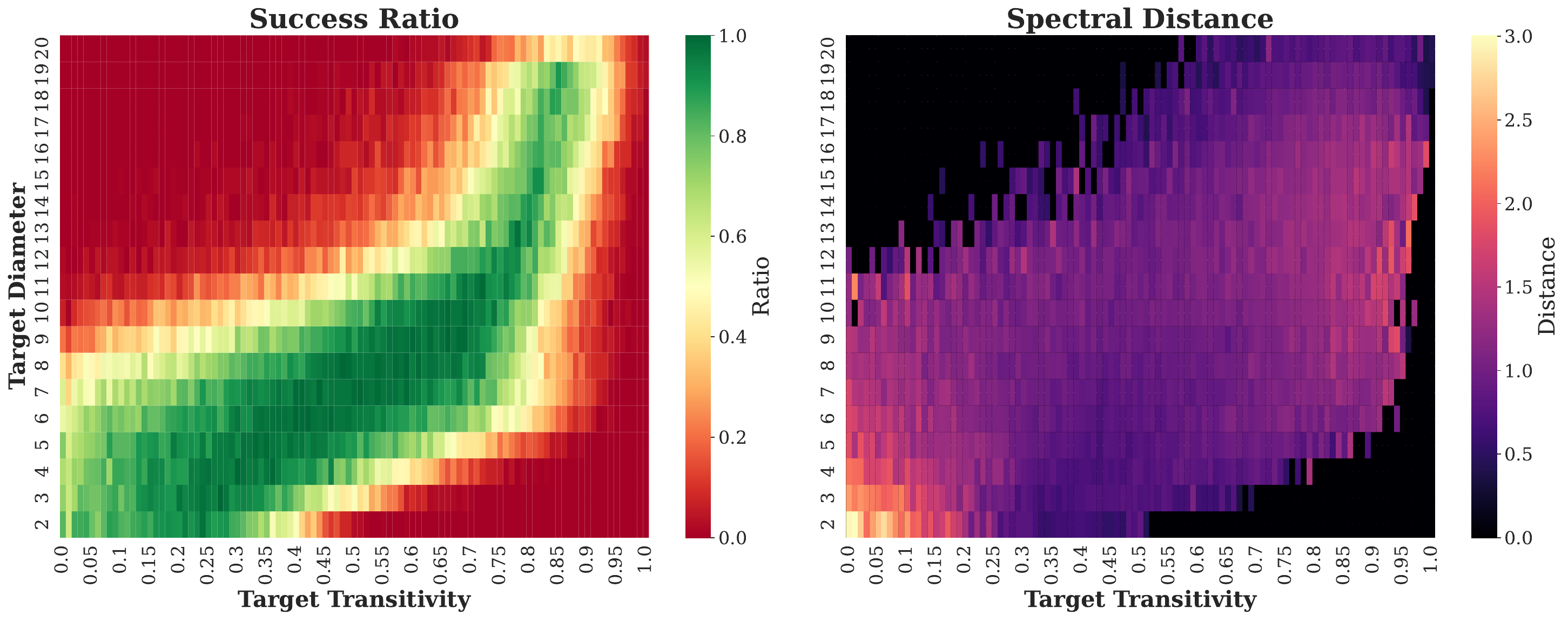}
	\caption{\label{fig:example_heatmap} Success ratio (left) and spectral distance (right) for edge density 0.2.}
	\includegraphics[width=1\columnwidth]{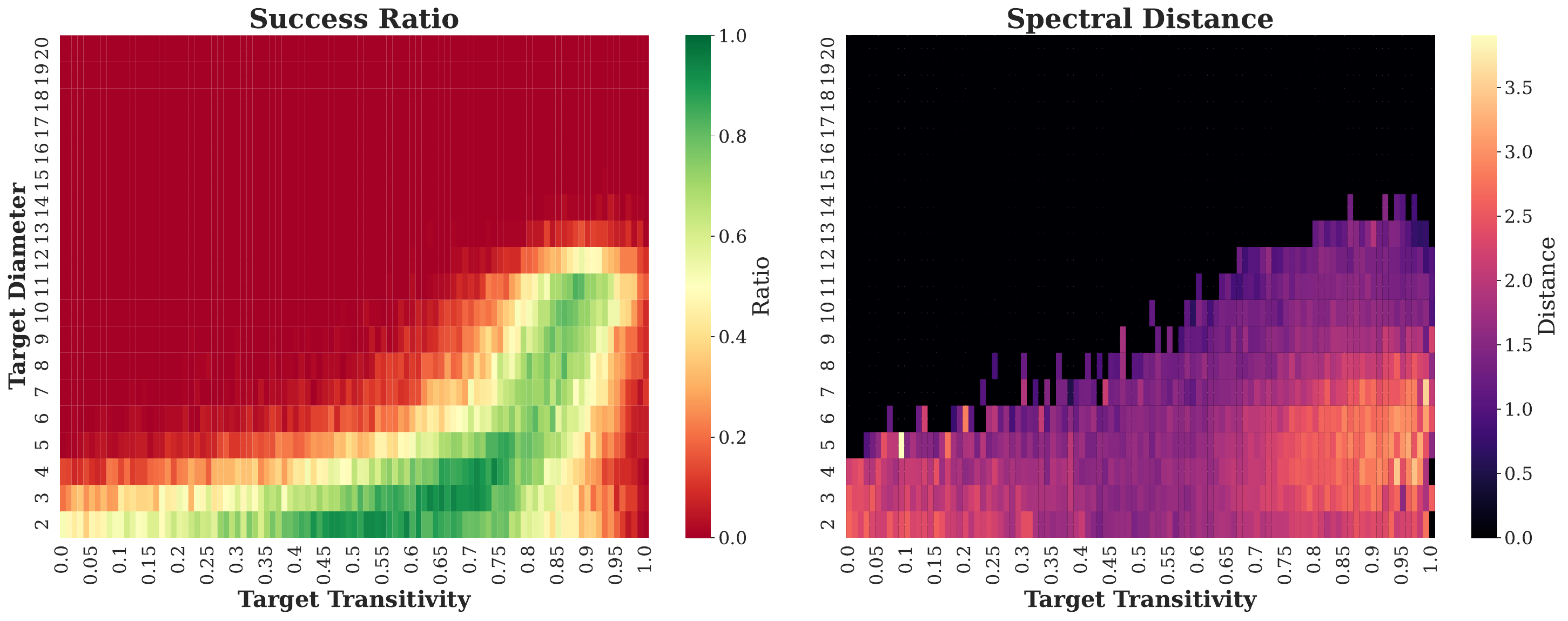}
	\caption{\label{fig:example_heatmap312} Success ratio (left) and spectral distance (right) for edge density 0.4. }
\end{figure}

Figures \ref{fig:example_heatmap} and \ref{fig:example_heatmap312} illustrate the relationship between success ratio and constraint strictness, as well as the impact of edge density on the feasible parameter space. Comparing success ratios reveals a shift; configurations that are accessible at lower edge densities may become topologically impossible as edges are added. 

		\subsection{Difference between MCMC and Hybrid MCMC}\label{subsec:diff_mcmc}
Next, we demonstrate the practical potential of our proposed hybrid framework by testing both, pure MCMC and hybrid MCMC, and measuring the structural difference of the seed and the output graphs of the MCMC and the hybrid MCMC.

\begin{figure}[htbp]
	\centering
	\includegraphics[width=\linewidth]{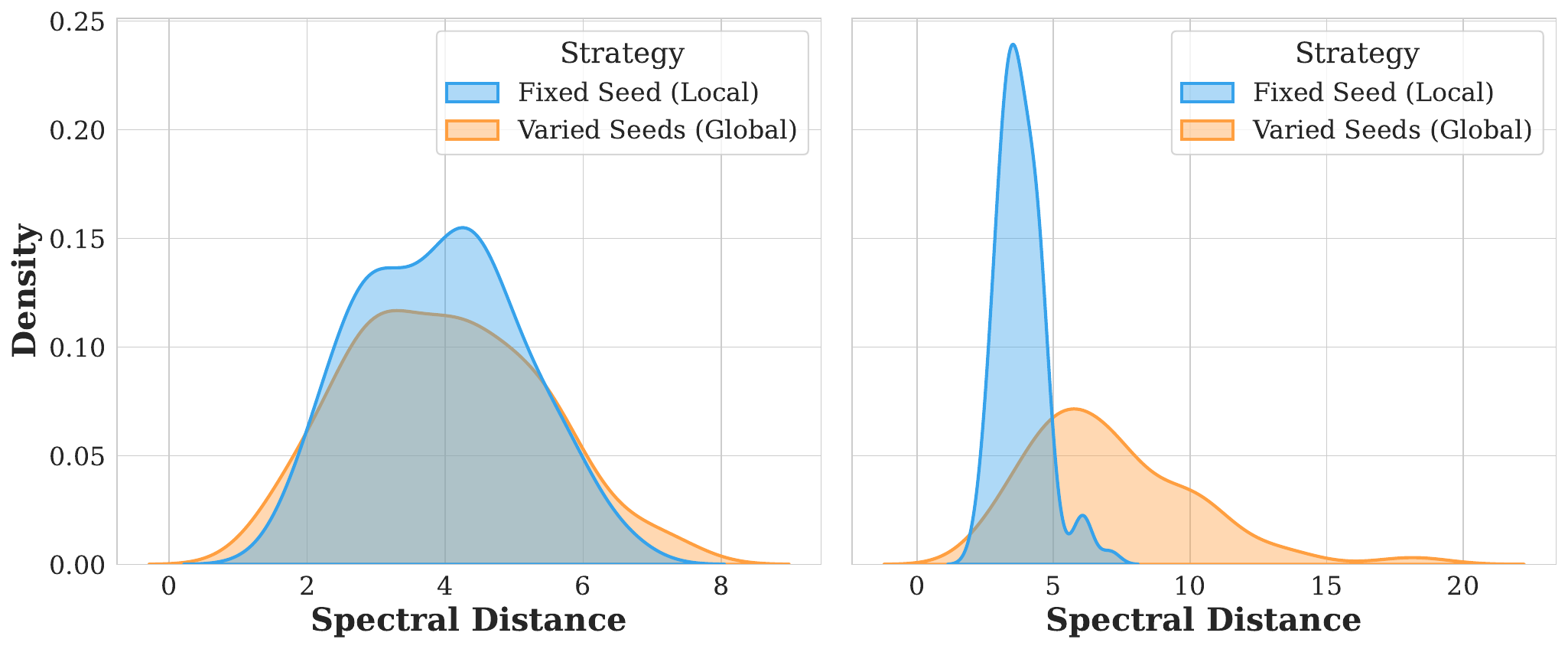}
	\caption{\label{fig:method_comp} Estimated density functions of the spectral distance (drift from seed). 
		\textbf{Left:} ($n=40$, $m=78$, $\diam=12$, $\cc=0.35$) Both methods find structurally similar solutions. 
		\textbf{Right:} ($n=40$, $m=195$, $\diam=4$, $\cc=0.4$) A different set of constraints allowing for a richer solution space. The standard MCMC (blue) remains clustered around the seed structure, while the hybrid method (orange) reveals the true structural diversity of the feasible set.}
\end{figure}

As shown in Figure \ref{fig:method_comp} (\textbf{Left}), when the feasible space is heavily constrained, both methods converge to similar structures and their distributions overlap, indicating that standard MCMC is sufficient when the target topology is unique. However, in the second case (\textbf{Right})---where the constraints allow for a rich variety of valid graphs—a clear divergence appears. While both methods find valid solutions, the standard MCMC remains trapped in the local neighborhood of its initialization (narrow peak), producing graphs structurally similar to the seed. In contrast, the hybrid approach successfully samples from the broader solution space, capturing a much richer set of graphs.

		\section{Discussion}
		
		In this work, we developed and demonstrated the practical utility of an efficient graph generation framework capable of strictly constraining clustering coefficient and diameter. While this framework serves as a practical tool for generating valid graphs under certain constraints, there are open questions that can be further investigated. These questions include, but are not limited to the following:
		\subsection{MCMC Ergodicity} Is it possible to theoretically characterize graphs that can be reached from a seed doing edge swaps, while maintaining the constraints for clustering and the diameter?  If the theoretic description is beyond reach is there a way to practically overcome the limitation and adapt an MCMC technique to sample the space of valid graphs, by overcoming ergodicity breaking?
		\subsection{Interplay of Local and Global Constraints}
		How do competing constraints on local transitivity (clustering) and global path lengths (diameter) interact to shape the overall topology of the graph? Understanding the structural boundaries imposed by these simultaneous constraints remains a theoretical challenge, as raising clustering decreases diameter by introducing shortcuts to the graph.

		Addressing these questions will bridge the gap between algorithmic graph generation and the theoretical understanding of structural phase transitions in complex networks.

		\begin{credits}
	\subsubsection{\ackname} This work was supported in part by the Dutch Research Council (NWO) Talent Programme ENW-Vidi 2021 under grant number VI.Vidi.213.163 (DF). This work used the Dutch national e-infrastructure with the support of the SURF Cooperative using grant no. EINF-15302
	
	\subsubsection{\discintname}
	The authors have no competing interests. 
	\end{credits}
	\bibliographystyle{splncs04}

	\bibliography{manuscript}

		\appendix
		\section{Metropolis-Hastings Algorithm}\label{app:MHalg}
		The Metropolis-Hastings (MH) algorithm is a Markov-chain Monte Carlo (MCMC) method used to generate a sequence of random samples from a probability distribution $\pi(G)$ that is difficult to sample from directly. In the context of graph generation, $\pi(G)$ typically represents the uniform distribution over the set of all valid graphs satisfying specific constraints.
		
		The algorithm constructs a Markov chain that asymptotically converges to the unique stationary distribution $\pi$. Let $G_t$ be the state of the graph at time step $t$. The transition to state $G_{t+1}$ proceeds as follows:
		
		\begin{enumerate}
			\item \textbf{Proposal:} A candidate graph $G'$ is generated from the current state $G_t$ using a proposal distribution $q(G' | G_t)$. In our case, this corresponds to the single-edge relocation or double-edge swap operation.
			
			\item \textbf{Acceptance Probability:} The candidate $G'$ is accepted with probability $A(G_t \to G')$, defined by the Metropolis-Hastings criterion:
			\begin{equation}
				A(G_t \to G') = \min \left( 1, \frac{\pi(G') q(G_t | G')}{\pi(G_t) q(G' | G_t)} \right).
			\end{equation}
			
			\item \textbf{Update:} Generate a uniform random number $u \in [0, 1]$.
			\begin{itemize}
				\item If $u \leq A(G_t \to G')$, accept the move: $G_{t+1} = G'$.
				\item Otherwise, reject the move and remain in the current state: $G_{t+1} = G_t$.
			\end{itemize}
		\end{enumerate}
		
		\subsection{Symmetric Proposal Simplification}
		In the case of simple graph moves (such as edge swaps), the proposal distribution is often symmetric, meaning the probability of proposing $G'$ from $G_t$ is equal to proposing $G_t$ from $G'$ ($q(G' | G_t) = q(G_t | G')$). Furthermore, for uniform sampling, the target distribution $\pi(G)$ is uniform ($\pi(G) \propto 1$ if valid, $0$ otherwise). 
		
		Under these conditions, the acceptance probability simplifies to a binary check:
		\begin{equation}
			A(G_t \to G') = 
			\begin{cases} 
				1 & \text{if } G' \text{ satisfies all constraints,} \\
				0 & \text{otherwise.}
			\end{cases}
		\end{equation}
		
		\subsection{Ergodicity}
		For the defined Markov chain to converge to the unique stationary distribution it is enough to show the following two properties: the Markov chain has to be irreducible ( every valid graph is constructable) and aperiodic (Certain graphs are not created periodically). If a Markov chain holds these properties it is said to be ergodic. 
		
		For aperiodicity it is enough to show that with a positive probability the new state is the same as the old state. For a chain's irreducibility one has to show that any valid state is reachable from every other valid state.
		
		In our graph generation case aperiodicity can be shown easily. It is enough to show that there exists a valid graph that can be turned into an invalid one by moving a single edge, as such a move would show that with a positive probability two consecutive states of the chain are identical.
		
		For irreducibility one would need to construct a set of valid moves between any two valid graphs (i.e. a set of edge swaps that connect to valid states).

	\end{document}